\documentclass[twocolumn,superscriptaddress,
 aps,
 prb]{revtex4-2}

\usepackage{graphicx}
\usepackage{dcolumn}
\usepackage{bm}
\usepackage{mathptmx}
\usepackage {hyperref}
\hypersetup{
     colorlinks   = true,
     citecolor    = blue,
     urlcolor=blue,
     linkcolor=blue
}
\usepackage{tabularx}
\usepackage{xcolor}
\usepackage[font=footnotesize,labelfont=bf]{caption}
\usepackage{siunitx}
\usepackage{csquotes}
\usepackage{amsmath}
\usepackage{chemformula}

\newcommand{\tc}{$T_\mathrm{c}$~}

\begin{document}

\title{Aluminium substituted yttrium iron garnet thin films with reduced Curie temperature}
\author{D. Scheffler}
\affiliation{Institute of Solid State and Materials Physics, Technische Universitaet Dresden, 01062 Dresden, Germany}
\author{O. Steuer}
\affiliation{Institute of Materials Science, Technische Universitaet Dresden,  01069 Dresden, Germany}
\affiliation{Institute of Ion Beam Physics and Materials Research, Helmholtz Zentrum Dresden-Rossendorf, 01328 Dresden, Germany}
\author{S. Zhou}
\affiliation{Institute of Ion Beam Physics and Materials Research, Helmholtz Zentrum Dresden-Rossendorf, 01328 Dresden, Germany}
\author{L. Siegl}
\author{S.T.B. Goennenwein}%
\author{M. Lammel}%
\affiliation{Department of Physics, University of Konstanz, 78457 Konstanz, Germany}

\date{\today}

\begin{abstract}

Magnetic garnets such as yttrium iron garnet (Y$_3$Fe$_5$O$_{12}$, YIG) are widely used in spintronic and magnonic devices. Their magnetic and magneto-optical properties can be modified over a wide range by tailoring their chemical composition. Here, we report the successful growth of Al-substituted yttrium iron garnet (YAlIG) thin films via radio frequency sputtering in combination with an ex situ annealing step. Upon selecting appropriate process parameters, we obtain highly crystalline YAlIG films with different Al$^{3+}$ substitution levels on both, single crystalline Y$_3$Al$_5$O$_{12}$ (YAG) and Gd$_3$Ga$_5$O$_{12}$ (GGG) substrates. With increasing Al$^{3+}$ substitution levels, we observe a reduction of the saturation magnetisation as well as a systematic decrease of the magnetic ordering temperature to values well below room temperature. YAlIG thin films thus provide an interesting material platform for spintronic and magnonic experiments in different magnetic phases.

\end{abstract}

\maketitle

\section{\label{sec:Intro}Introduction}

In the field of insulator spintronics, magnetic materials are usually used in their ordered phase, i.e. well below their respective magnetic ordering temperature \cite{Weiler2012,Nakayama2013,Althammer2013,Cornelissen2015,Serga2010,Chumak2015,Goennenwein2015,Meyer2017,Ganzhorn2016}. Therefore, also almost all spin transport experiments have been performed in the ferro-, ferri- or antiferromagnetic phase \cite{Weiler2012,Nakayama2013,Althammer2013,Cornelissen2015,Serga2010,Chumak2015,Goennenwein2015,Meyer2017,Ganzhorn2016}. Moreover, a few attempts to study spin transport in the paramagnetic phase have also been made, either by using paramagnetic materials \cite{Lammel2019,Oyanagi2019,Oyanagi2021}, or by performing the experiments at elevated temperatures above the ordering temperature \cite{Schlitz2021,Aqeel2015,Martin-Rio2022}. A comprehensive understanding of spin transport in the paramagnetic phase nevertheless is lacking, not to mention its evolution across the magnetic phase transition. This makes systematic experiments across the magnetic phase transition, i.e. from the ferromagnetic to the paramagnetic phase, highly desirable to establish a robust understanding of spin transfer as well as magnon generation and propagation processes. In addition, magnetic fluctuations are enhanced around the phase transition, allowing to study the impact of such fluctuations on spin transport. 

The prototype material for spin transport studies \cite{Nakayama2013,Althammer2013,Cornelissen2015,Serga2010,Chumak2015,Goennenwein2015,Schlitz2021,Uchida2015} is the magnetic insulator yttrium iron garnet Y$_3$Fe$_5$O$_{12}$ (YIG), as it has a very low Gilbert damping parameter \cite{Onbasli2014,Schmidt2020,Hauser2016,Dubs2020} and a small coercive field \cite{Onbasli2014,Dubs2020,Krockenberger2008}. This makes YIG an ideal material for magnon transport experiments and spin Hall effect driven spin transport studies \cite{Nakayama2013,Althammer2013,Cornelissen2015,Serga2010,Chumak2015,Goennenwein2015,Schlitz2021,Uchida2015}. However, YIG has proven problematic for spin transport experiments across the magnetic phase transition due to its relatively high Curie temperature of $T_\mathrm{c} = 559$\,K \cite{Anderson1964}. Experiments at temperatures close to or above \tc are hampered by a finite thermally induced electrical conductivity of YIG \cite{Thiery2018}. In addition, a significant interdiffusion of Pt in YIG/Pt heterostructures has been reported for $T>470$\,K, which leads to a significant deterioration of the interface and the magnetic properties of YIG \cite{Schlitz2021}.
 
One way to circumvent such high temperature issues is to lower $T_\mathrm{c}$. As the magnetic properties of YIG are defined by the Fe$^{3+}$ ions, a lower \tc can be achieved by substituting the Fe$^{3+}$ ions with non-magnetic ions. This has been successfully accomplished almost exclusively for bulk crystals by substituting with, amongst others, Al$^{3+}$ ions \cite{Harrison1961,Geller1964,Dionne1970,Roeschmann1981,Grasset2001,Ravi2007,Motlagh2009,Chen2005,Grasset2001}. By increasing the Al$^{3+}$ substitution, both Curie temperature \tc and saturation magnetisation $M_\mathrm{s}$ are reduced. Sufficiently high substitution levels $x$ allow to tailor \tc of the resulting yttrium aluminium iron garnet (Y$_3$Al$_x$Fe$_{5-x}$O$_{12}$, YAlIG) from the value of pure YIG down to values well below room temperature. It has been shown that the desired magnetic properties of YIG such as low magnetic damping and low coercivity are conserved upon substitution with Al$^{3+}$ ions \cite{Harrison1961,Dionne1969}. However the studies of YAlIG are almost exclusively limited to bulk and powder materials \cite{Harrison1961,Geller1964,Dionne1970,Roeschmann1981,Grasset2001,Ravi2007,Motlagh2009,Chen2005,Grasset2001},  while thin films are required for typical spintronic devices. So far, only YAlIG thin films up to $x=0.783$ were investigated \cite{Liang2018}, which is not sufficient to reduce \tc down to room temperature. 

Here, we report the successful fabrication of YAlIG thin films with different Al$^{3+}$ substitution levels $1.5 \leq x \leq 2$ by radio-frequency (RF) sputtering deposition and a subsequent annealing step. We evaluate the influence of the Al$^{3+}$ substitution level on the structural and magnetic properties of our thin films. To this end, we study the impact of different substrate materials and annealing temperatures on the structural and magnetic quality. X-ray diffraction (XRD) is utilised to show that our sputtered and post-annealed YAlIG thin films have a high crystal quality. SQUID vibrating sample magnetometry demonstrates the successful reduction of \tc to values well below room temperature, as well as coercive fields comparable to pure YIG films. We utilise broadband ferromagnetic resonance measurements to investigate the dynamic magnetic properties and the magnetic anisotropy of our YAlIG thin films.

\begin{table*}
\caption{\label{tab:data} Summary of nominal Al$^{3+}$ substitution level $x$, annealing temperature $T_A$, thickness $t$, lattice parameters $a$ (length) and $\alpha$ (angle), saturation magnetisation $M_\mathrm{s}$ at 20\,K, Curie temperature \tc and the upper limit for the Curie temperature $T_\mathrm{c,max}$ for the Y$_3$Al$_x$Fe$_{5-x}$O$_{12}$ films.}
\begin{tabular}{c@{\hskip 0.1in}c@{\hskip 0.1in}c@{\hskip 0.1in}c@{\hskip 0.1in}c@{\hskip 0.1in}c@{\hskip 0.1in}c@{\hskip 0.1in}c@{\hskip 0.1in}c@{\hskip 0.1in}c}
\hline
\hline
substrate & $x$ & $T_A$ (°C) & $t$ (nm) & $a_\mathrm{YAlIG}$ (nm) & $\alpha$ (deg)& $M_\mathrm{s} @$\,20\,K (kAm$^{-1}$) & \tc (K) & $T_{c,max}$ (K)\\
\hline
YAG & 1.5  & as-dep.  & 50 & amorphous  & amorphous & $<5$ & -- & --\\
YAG & 1.5  & 700      & 45 & 1.2276  & 90     &  41   & 270 & 300\\
YAG & 1.5  & 800      & 45 & 1.2276  & 90     &  41   & 262 & 300\\
YAG & 1.75 & 800      & 37 & 1.2266  & 90     &  16   & 156 & 190\\
YAG & 2    & 800      & 45 & 1.2205  & 90     &  10   & 102 & 140\\
 \hline
GGG  & 1.5 & as-dep.  & 50 & amorphous & amorphous & -- & -- & --\\
GGG  & 1.5 & 700      & 45 & 1.2355 & 90.46  & -- & -- & --\\
GGG  & 1.5 & 800      & 45 & 1.2348 & 90.49  & -- &  -- & --\\

\hline
\hline
\end{tabular}
\end{table*}

\section{\label{sec:methods}Experimental Methods}

The YAlIG films were grown on commercially available (111)-oriented single crystalline yttrium aluminium garnet (\ch{Y3Al5O12},YAG, cubic lattice parameter $a_\mathrm{YAG}=1.201$\,nm) and gadolinium gallium garnet (\ch{Gd3Ga5O12}, GGG, $a_\mathrm{GGG}=1.240$\,nm) substrates by RF-sputtering in a UHV system (base pressure below  $1\times10^{-8}$\,mbar). Three different 2 inch diameter targets with Al$^{3+}$ substitution levels of $x=1.5,1.75$ and 2 were used. These substitution levels were selected to obtain thin films with a Curie temperature \tc close to or below room temperature based on the existing literature for bulk YAlIG \cite{Geller1964,Grasset2001,Chen2005}.

Prior to deposition, the substrates were cleaned subsequently in acetone, isopropanol and distilled water in a ultrasonic bath for 10 min each. After transferring the substrates into the sputtering chamber, they were annealed in situ at 200\,°C for 1\,h in vacuum to eliminate residual adsorbed water. The YAlIG films were deposited at room temperature. The thermal energy for the crystallisation of the YAlIG thin films was provided by an ex situ post-annealing in a two-zone furnace. The detailed deposition and annealing parameters are summarised in table \ref{tab:data_exp}. Temperatures of 700\,°C and above were reported to be sufficient for a complete crystallisation of YIG thin films of comparable thickness \cite{Ding2020a,Lammel2022,Chang2014}. The partial oxygen pressure was applied to counteract oxygen losses of the films during the annealing \cite{Lammel2022}. All samples examined in this study are summarised in table \ref{tab:data}. 

\begin{table}[h]
\caption{\label{tab:data_exp} Summary of the deposition and annealing parameters for the fabrication of the YAlIG thin films.}
\begin{tabular}{l@{\hskip 0.2in}l@{\hskip 0.2in}l}
\hline
\hline
sputter deposition 	&	target-substrate distance	& 206\,mm\\
					&	sample holder rotation rate 	& 5\,rpm\\
					&	pressure (Ar)				& $4.3 \times 10^{-3}$\,mbar\\
					&  	Ar flow rate					& 20\,sccm\\
					&	sputtering power				& 75\,W\\
\hline
annealing			&	heating rate					& 15 K/min\\
					&	pressure	 (O$_2$)				& 3 mbar\\
					&	annealing temperature		& $700-800$\,°C\\
					&	annealing time 				& 240 min\\
\hline
\hline
\end{tabular}
\end{table}

Various X-ray diffraction measurements (symmetrical $2\theta$-$\omega$ scan, $\omega$-scan, reciprocal space mapping) were performed using Cu-$K_\mathrm{\alpha1}$ radiation\,($\lambda=0.15406\,$nm) to evaluate the structural properties of the film.
X-ray reflectivity was used to measure the film thickness. 

Rutherford backscattering spectroscopy random (RBS/R) experiments at a backscattering angle of 170° were performed to investigate the chemical composition of selected YAlIG films, using a 1.7\,keV He$^+$ ion beam with a diameter of about 1\,mm generated by a 2\,MV van-de-Graaff accelerator. The backscattering angle was 170°. The measured data was fitted to a calculated spectra using the \textit{SIMNRA} software \cite{Mayer1997}. Due to the low mass of oxygen, only the elemental ratios between Fe, Al and Y are used for quantitative analysis of the chemical composition.

The static magnetic properties were investigated by superconducting quantum interference device vibrating sample magnetometry (SQUID VSM). For an external magnetic field orientation within the sample plane during the measurement (in-plane, ip), the sample was mounted onto a quartz rod; for an orientation of the external magnetic field parallel to the surface normal (out-of-plane, oop) the sample was mounted between plastic straws. The magnetic moment of the sample was recorded in dependence of the external field (maximum field $\mu_\mathrm{0}H_\mathrm{max}=\pm$\,3\,T, minimal field steps of $\Delta\mu_\mathrm{0}H=2$\,mT) and the temperature (2\,K to 360\,K, temperature steps $\Delta T=2$\,K). Note that due to the large paramagnetic background signal of the GGG substrate and the correspondingly larger background subtraction error, we focused on the YAlIG thin films on YAG substrates for the magnetisation characterization. For all magnetometry data shown in this work, the magnetic contributions of the YAG substrate were subtracted. For temperature dependent measurements, this includes the diamagnetic contribution and an additional paramagnetic contribution from contaminations within the YAG substrate, which combined where fitted to $m_\mathrm{sub} =a/(T+b) +c$ with $a,b$ and $c$ as fitting parameters. For field dependent measurements, the substrate contribution is defined by the linear slope at high fields. 

Broadband ferromagnetic resonance (bb-FMR) measurements  were performed to investigate the dynamic magnetic properties and the magnetic anisotropy of two representative samples, both with similar nominal composition and annealing temperature($x=1.5$, $T_A=800$\,°C), but on different substrates. Therefore, the samples are placed on a coplanar waveguide (CPW) with the thin film facing towards the center conductor of the CPW. The CPW is connected to two ports of a vector network analyser (VNA, PNA N5225B from Keysight). To control the temperature of the sample and to apply an external magnetic field, the CPW assembly was inserted into a 3D vector magnet cryostat. Using the VNA, we record the complex microwave transmission parameter $S_\mathrm{21}$ as a function of frequency $f=1-30$\,GHz at several fixed static magnetic fields $H$. The magnetic field was applied either parallel (ip) or perpendicular (oop) to the sample plane. The measured $S_{21}$ was corrected for the frequency dependent and magnetic field dependent background (see moving field reference method in \cite{Maier-Flaig2018}) and fitted by a complex Lorentzian function to extract the FMR resonance frequency $f_\mathrm{res}$ and linewidth $\Delta f$.

\section{\label{sec:results}Results}
\subsection*{Structural characterisation}

\begin{figure}[h!]
\includegraphics{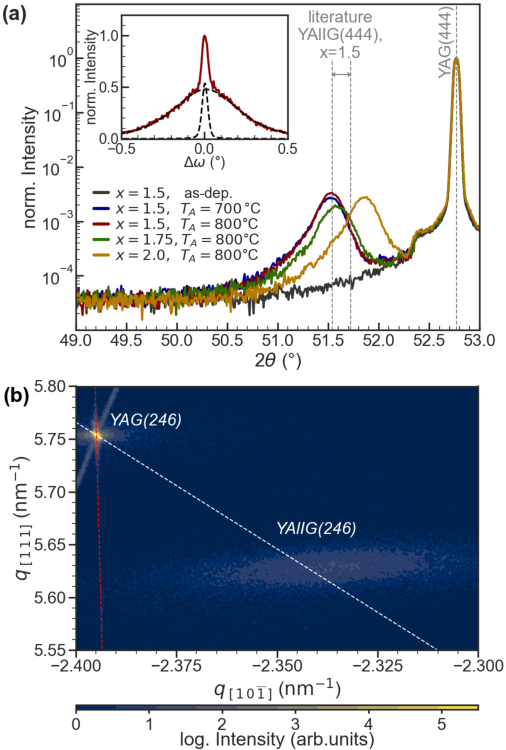}
\caption{\label{fig:XRD-YAG} XRD-analysis of YAG/YAlIG films: \textbf{(a)} Symmetric $2\theta$-$\omega$ scan for different annealing conditions and compositions. The two grey dashed lines denote the range of values reported in literature for $x=1.5$ \cite{Gilleo1958b,Geller1964,Motlagh2009,Musa2017,Rodic2001,Bhalekar2019}. The inset shows the $\omega$ scan of the YAlIG film annealed at 800°C, $x=1.5$. The dashed lines indicate the numerical fit of the data to two independent Pseudo-Voigt functions. \textbf{(b)} RSM of the asymmetric (246)-peak of a sample annealed at 800°C, $x=1.5$. The white dashed line indicates the peak position for a growth of fully relaxed YAlIG and the red dashed lines indicates the peak position for a growth of fully strained YAlIG.}
\end{figure}

X-ray diffraction measurements carried out on YAlIG thin films with different compositions revealed a high crystalline quality of the respective YAlIG thin films after the annealing step. 
The symmetrical $2\theta$-$\omega$ XRD scans of YAlIG thin films with different substitution levels grown on YAG are shown in Figure~\ref{fig:XRD-YAG}a. No diffraction peak that can be ascribed to the YAlIG thin films is observed prior to annealing, supporting the notion that the thermal energy during deposition is not sufficient for crystallisation. The necessary thermal energy is provided by the subsequent annealing step, which leads to the crystallisation of the YAlIG layer. For all chemical compositions the crystalline nature of the annealed thin films is validated by the diffraction peak which corresponds to YAlIG\,(444) planes.
For substituted yttrium iron garnet, the lattice parameter is changing according to the size of the substitute ion \cite{Gilleo1958b}. As Al$^{3+}$ is smaller than Fe$^{3+}$, the lattice parameters of the YAlIG thin films (cp. table \ref{tab:data}) are reduced compared to YIG ($a_\mathrm{YIG}=1.2376$\,nm \cite{Gilleo1958b}), as expected. Likewise the lattice parameter is decreasing towards higher substitution levels, which becomes apparent by the shift towards higher diffraction angles. The lattice parameter falls well within the range given by the literature ($1.2235$\,nm$\leq a_\mathrm{YAlIG} \leq 1.2276$\,nm) \cite{Gilleo1958b,Geller1964,Motlagh2009,Musa2017,Rodic2001,Bhalekar2019}, which is illustrated by the grey dashed lines in Fig.~\ref{fig:XRD-YAG}a for $x=1.5$. 
Comparing the XRD patterns does not indicate any influence of the annealing temperature on the maximum peak position of thin films with the same Al$^{3+}$ substitution level, aside from a small change in the peak intensity. This might indicate a more complete crystallisation at higher temperatures.

A slight peak asymmetry towards lower diffraction angles can be observed independently from the annealing condition or chemical composition which might originate from a partially strained film or a gradient of chemical composition due to diffusion of Al$^{3+}$ from the substrate into the film.
Since the lattice mismatch $\epsilon$ between the YAG substrate and our YAlIG thin film is relatively high ($\epsilon=\frac{a_\mathrm{YAlIG}-a_\mathrm{sub}}{a_ {sub}}= 2.23\%, 2.14\%, 1.63\%$ for $x=1.5,1.75$ and $2$)\cite{Gross2018}, we anticipate a relaxed growth for the given film thickness. This is confirmed by reciprocal space mapping (RSM, cp. Fig.~\ref{fig:XRD-YAG}b), as the asymmetric YAlIG (246) peak is located on the calculated line for a fully relaxed crystal. Although the lattice mismatch is reduced at higher substitution levels, no change of the described growth mechanism is found. Further structural investigation via RSM to explain the peak asymmetry observed in the $2\theta$-$\omega $ XRD scan are not possible due to the low intensity of the asymmetric (246) peak.  The $\omega$-scan of a YAlIG film annealed at $T_A=800$\,°C, given in the inset of Fig.~\ref{fig:XRD-YAG}a, shows a superposition of two diffraction peaks with different full width at half maximum (FWHM) values centred around the same diffraction angle. The FWHM of each peak was determined by numerically fitting of the data to the sum of two Pseudo-Voigt functions as $0.04^\circ$ and $0.46^\circ$. A similar behaviour has been reported for different heteroepitaxially relaxed thin film materials \cite{John2021,Becht1997,Webster2010,Durand2011,Kaganer2009}. The sharp peak is associated with areas of higher structural order, i.e. an epitaxially strained crystalline layer close to the substrate/film interface. Upon further crystallisation, the influence of the substrate decreases and the film continues to grow relaxed with the incorporation of defects like dislocations or a higher mosaicity, which is ascribed to the broad peak feature. As 91\% of the total peak area contributes to the broad peak, we conclude that the majority of the film features a higher degree of defects. 

\begin{figure}[ht]
\includegraphics{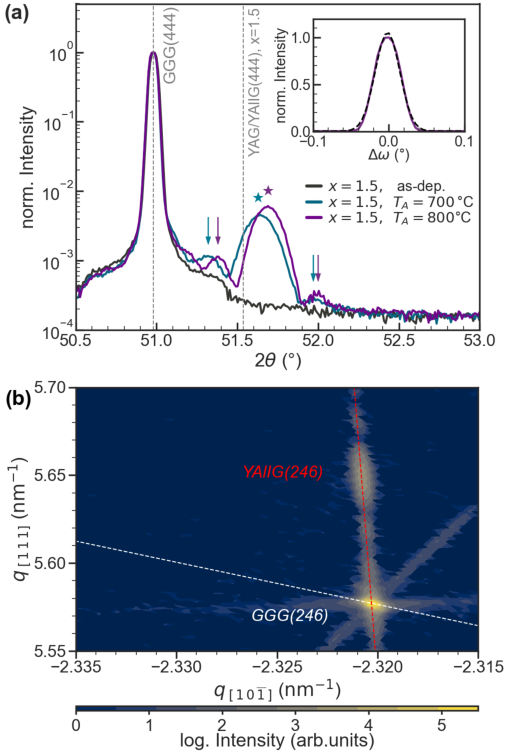}
\caption{\label{fig:XRD-GGG} XRD-analysis of GGG/YAlIG films with $x=1.5$: \textbf{(a)} Symmetric $2\theta$-$\omega$ scan of films annealed at different temperatures. The $\star$ indicate YAlIG\,(444) peak, the arrows indicate the first Laue oscillation of YAlIG. The inset shows the $\omega$ scan of the YAlIG film annealed at 800°C. The dashed line indicates the numerical fit of the data to a Pseudo-Voigt function. \textbf{(b)} RSM of the asymmetric (246)-peak of a sample annealed at 800°C, where the white dashed line indicates the peak position for a growth of fully relaxed YAlIG and the red dashed lines indicate the peak position for a growth of fully strained YAlIG.}
\end{figure}

To evaluate the effect of a smaller lattice mismatch, YAlIG samples with $x=1.5$ were also prepared on GGG substrates. This reduces the lattice mismatch down to $-1\%$. As for the samples grown on YAG, a post-annealing step is necessary for the crystallisation of YAlIG (444) (cp. Fig. \ref{fig:XRD-GGG}a). However, the peak position of YAlIG (444) in the symmetrical $2\theta$-$\omega$ XRD scan is shifted towards higher diffraction angles compared to the YAlIG samples grown on YAG, visualised by the grey dashed line. This reduction of the distance between the YAlIG (111) lattice planes indicates a fully strained growth of YAlIG on GGG due to the reduced lattice mismatch while keeping the film thickness at 45\,nm. This is confirmed by RSM analysis (cp. Fig.~\ref{fig:XRD-GGG}b), as the asymmetric YAlIG\,(246) diffraction peak is located on the calculated line for a fully strained film. Hence, YAlIG is strained in the film plane and compressed perpendicular to the film surface in comparison to cubic YAlIG, resulting in a rhomboidal YAlIG crystal. The lattice parameters based on the RSM are summarised in table~\ref{tab:data}.
 The high crystalline quality of the YAlIG thin films on GGG is confirmed by the very low FWHM of 0.04$^\circ$ in the $\omega$-scan (cp. Fig.~\ref{fig:XRD-GGG}a, inset), which is in the range of the resolution of the X-ray measuring device. In addition, the first Laue oscillation is visible in the symmetric $2\theta$-$\omega$ scan, indicating high structural ordering of the film. 
 The increase in lattice parameter $\alpha$, the angle between the unit cell vectors, suggests an enhancement of the strain for higher annealing temperatures. We attribute this to a change of lattice mismatch at higher temperatures due to different thermal coefficients of substrate and film, which was shown for different rare earth garnets \cite{Geller1969}.

\subsection*{Chemical composition}

\begin{figure}[h]
\centering
\includegraphics{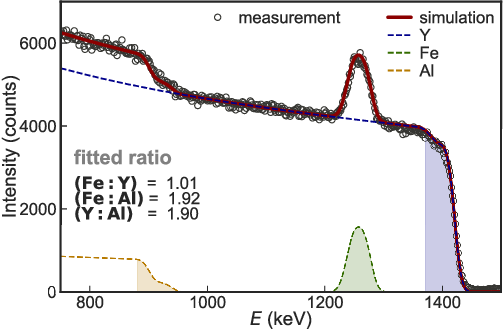}
\caption{\label{fig:RBS} RBS spectra for a YAlIG film with $x=1.5$ on YAG annealed at $T_A=800$\,°C. The shaded areas indicate the thin film contribution of the RBS spectra.}
\end{figure}

RBS measurements were performed for selected samples to investigate the chemical composition, revealing a significant deviation of the nominal composition. Here, we will present and discuss the RBS results exemplary for a YAlIG film ($x=1.5$, $T_{A}=800$\,°C) on YAG as given in Fig. \ref{fig:RBS}, however, similar correlations were found for all investigated samples.

 Going from high backscattered ion energy to lower energies of the spectrum, the elemental contributions of Y, Fe, Al and O (not shown in Fig. \ref{fig:RBS}) can be distinguished. The first approximately 80\,keV of each elemental contribution belong to the YAlIG thin film, indicated by the coloured areas in Fig. \ref{fig:RBS}. As both the film and the YAG substrate contain Y,Al and O, the respective contributions of the film are superimposed by high mass contributions originating from the substrate towards lower energies.
The simulated spectrum (cp. solid line in Fig. \ref{fig:RBS}) is in good agreement to the measurement and the elemental ratios are determined from the fit as $(\mathrm{Fe}:\mathrm{Y}) = 1.01$, $(\mathrm{Fe}:\mathrm{Al}) = 1.92$ and $(\mathrm{Y}:\mathrm{Al}) = 1.90$. These values differ from the values for the nominal stoichiometric composition of the target  for $x=1.5$, i.e.$(\mathrm{Fe}:\mathrm{Y})_\mathrm{target} = 1.17$, $(\mathrm{Fe}:\mathrm{Al})_\mathrm{target} = 2.33$ and $(\mathrm{Y}:\mathrm{Al})_\mathrm{target} = 2$. The most significant change is the decreased $(\mathrm{Fe}:\mathrm{Al})$ ratio, indicating a lowered Fe content and an increased Al content. This is confirmed by the decrease of the $(\mathrm{Fe}:\mathrm{Y})$ ratio and the $(\mathrm{Y}:\mathrm{Al})$ ratio. Therefore we conclude that our YAlIG films have a lower Fe$^{3+}$ content and an increased Al$^{3+}$ content compared to the chemical composition of the target, while there's no significant change of the Y$^{3+}$ content.
We attribute this difference to changes of the chemical composition during the sputtering process, where it was shown that the chemical composition can be altered by different process conditions \cite{Wu2018,Park2001}. This has the potential to enable fine-tuning of the chemical composition by adjusting the deposition process. 

\subsection*{Magnetometry}

\begin{figure}[h]
\centering
\includegraphics{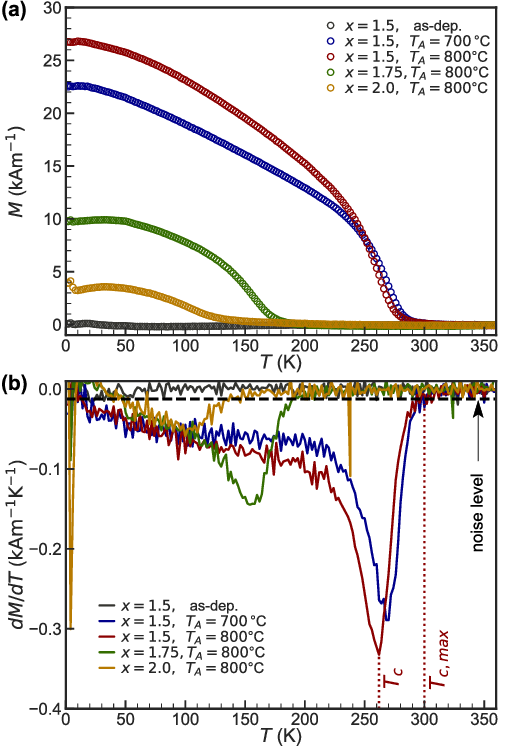}
\caption{\label{fig:magneticMT} \textbf{(a)} SQUID-VSM $M(T)$ measurements with a static field of $\mu_0H=0.1$\,T applied in the film plane. The Curie temperature $T_c$ was determined at the minimum of the first derivative $\frac{dM}{dT}$ \textbf{(b)}. To get an upper limit of the Curie temperature, $T_\mathrm{c,max}$ was determined at the temperature were the first derivative is lower as the noise level, which is indicated as a dashed black line.}
\end{figure}

The SQUID VSM measurements demonstrate ferromagnetism with a reduced Curie temperature that scales with the Al$^{3+}$ substitution level for YAlIG thin films grown on YAG substrates.
 The temperature dependent magnetisation is given in Fig.~\ref{fig:magneticMT}a. For the as-deposited sample, no net magnetisation was observed in the M(T) curve. Upon annealing and crystallising into YAlIG a magnetic phase transition emerges. The Curie temperature for all substitution levels is substantially lowered with respect to that of pure YIG ($T_\mathrm{c}=599$\,K) \cite{Anderson1964}, as expected by theory and previous studies in YAlIG \cite{Roeschmann1981,Geller1964,Motlagh2009}, since the ferrimagnetic coupling between the Fe$^{3+}$ ion sublattices is weakened by the substitution of Fe$^{3+}$ ions with non-magnetic Al$^{3+}$ ions. Consequentially, \tc decreases for higher substitution levels which is also represented in our data.
Experimentally, the ferrimagnetic-paramagnetic transition at \tc is broadened by the presence of the static magnetic field of $\mu_0H=0.1$\,T, by non-uniform temperature sweep rates and by inhomogeneities of composition within the sample. Therefore, \tc was determined at the minimum of the first derivative of the magnetisation with respect to the temperature$\frac{dM}{dT}$ (cp. Fig. \ref{fig:magneticMT}b). In addition, we define the upper limit for the Curie temperature $T_\mathrm{c,max}$ at the first temperature where $\frac{dM}{dT}$ is below the noise level of the measurement. Both, \tc and $T_\mathrm{c,max}$ for all YAG/YAlIG samples are summarised in Table~\ref{tab:data}.

\begin{figure}[h]
\centering
\includegraphics[width=0.5\textwidth]{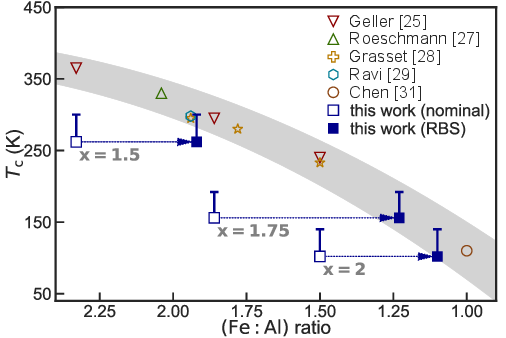}
\caption{\label{fig:Tc-x} Curie temperature \tc of YAlIG with different Fe:Al ratios, reported in the literature \cite{Roeschmann1981,Geller1964,Gilleo1958b,Ravi2007,Grasset2001,Chen2005} and the results of this study. The shaded area is a guide for the eye for the expected range of \tc based on the literature. For the results of this study, the errorbars indicate the upper limit $T_\mathrm{c,max}$ and the arrow indicates the shift due to the difference between the nominal $(\mathrm{Fe}:\mathrm{Al})_\mathrm{target}$ ratio and the $(\mathrm{Fe}:\mathrm{Al})$ ratio measured via RBS.}
\end{figure}

 Regarding the nominal ratio $(\mathrm{Fe}:\mathrm{Al})_\mathrm{target}$, the measured \tc of our YAlIG thin films is significantly lower than that reported in the literature \cite{Geller1964,Dionne1970,Roeschmann1981,Grasset2001,Ravi2007,Chen2005}, as displayed in Fig. \ref{fig:Tc-x}. However, taking into account the actual $(\mathrm{Fe}:\mathrm{Al})$ ratio of the YAlIG thin film, as measured by RBS,  and the maximum Curie temperature $T_\mathrm{c,max}$, the measured \tc of our YAlIG thin films is consistent with the literature, as indicated by the arrows and errorbars in Fig \ref{fig:Tc-x}.
In addition to the chemical composition, other effects might cause a reduction of the Curie temperature. For bulk YAlIG, the distribution of Fe$^{3+}$/Al$^{3+}$ among the a-site and d-site within the garnet structure could be altered by different annealing conditions, which leads to a change of Curie temperature and saturation magnetisation \cite{Roeschmann1981}. While the latter is strongly influenced by the changed distribution, only small changes for the Curie temperature of below 10\,K are reported. Since our rare earth garnet thin films feature thicknesses $t\geq 37\,$nm, dimensionality effects on the Curie temperature should not be relevant \cite{Chern1997,Takeuchi1983}.

\begin{figure}
\centering
\includegraphics{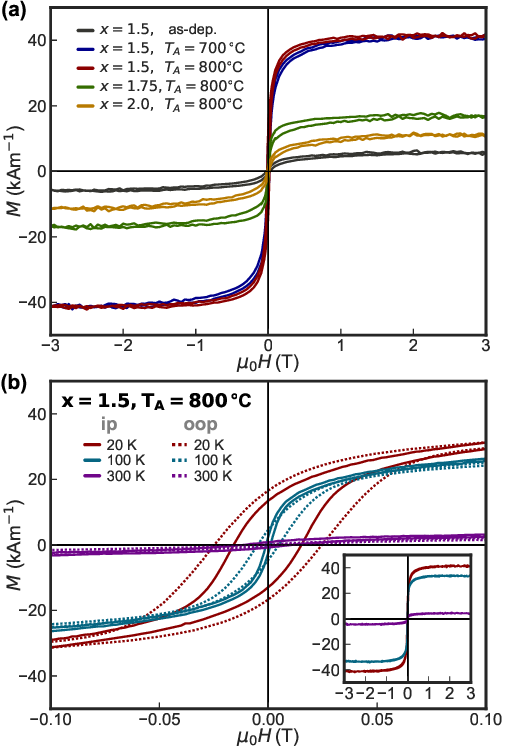}
\caption{\label{fig:magnetic} SQUID-VSM $M(H)$ hysteresis loops of YAlIG thin films grown on YAG. \textbf{(a)} In-plane $M(H)$ hysteresis loops up to $\mu_\mathrm{0}H=\pm 3\,$T at 20 K for different annealing temperatures and compositions. \textbf{(b)} In-plane $M(H)$ hysteresis loops at different temperatures for fields up to $\mu_\mathrm{0}H=\pm 0.1\,$T for a sample with $x=1.5$ annealed at $T_A=800$\,°C. The inset shows the complete field range of $\mu_\mathrm{0}H=\pm 3\,$T.}
\end{figure}

The magnetisation in dependence of the externally applied  magnetic field at 20\,K for different Al$^{3+}$ substitution levels $x$ is given in Fig.~\ref{fig:magnetic}a. The observed hysteresis loop further corroborates the ferrimagnetic ordering  of the YAlIG film at low temperatures. Above 1\,T the YAlIG thin films are saturated. The saturation magnetisation $M_\mathrm{s}$ for each sample is summarised in Table~\ref{tab:data}. As already implied by the M(T) curves, the field dependent measurements show a decreasing $M_\mathrm{s}$ towards higher substitution levels. This is explained by the predominant substitution of Fe$^{3+}$ on the tetrahedral sites in YAlIG \cite{Geller1964}.
Please note that the amorphous, as-deposited sample shows a hysteresis with $M_s=5$\,kAm$^{-1}$ although no sign of ferrimagnetism was observed in the temperature dependent measurements (cp. Fig.~\ref{fig:magneticMT}a). We attribute this apparent absence of ferrimagnetic order in the M(T) measurements to the subtraction of the paramagnetic background, which is particularly delicate for samples with an inherently small ferrimagnetic magnetisation. The ferrimagnetism of the as-deposited sample can be explained by iron rich areas within the film which can couple ferrimagnetically. However, as the as-deposited YAlIG thin films exhibit no signs of a crystalline ordering, the ferrimagnetism is expected to be weak compared to the annealed and crystallised YAlIG samples, which is corroborated by the temperature as well as the field dependence of the magnetisation (cp. Fig. \ref{fig:magneticMT} and \ref{fig:magnetic}).

To extract the coercivity of the YAlIG thin films, the magnetic hysteresis curves at different temperatures and sample orientations at low fields $\mu_\mathrm{0}H < 100$\,mT are representatively shown in Fig.~\ref{fig:magnetic}b for a YAlIG thin film with $x=1.5$. At 300\,K, almost no coercivity and magnetisation can be measured as the sample is in vicinity of the Curie temperature ($T_\mathrm{c}=262$\,K, $T_\mathrm{c,max}=300$\,K). As reported for rare earth garnet thin films, the coercive field of our YAlIG thin films increases with lower temperatures \cite{Vertesy1995,Mitra2017}. For an external magnetic field applied in the sample plane, the coercive field of $\mu_\mathrm{0}H_\mathrm{c}=15$\,mT at 20\,K and $\mu_\mathrm{0}H_\mathrm{c}=1$\,mT at 100\,K are in the order of the coercive fields of yttrium iron garnet thin films grown by magnetron sputtering on YAG \cite{Mitra2017}. 
No square loop is observed for either field direction in our YAlIG thin films. The X-ray analysis suggests the existence of crystal defects, which might act as pinning centres for the domain wall motion. However, the absence of a distinct magnetic hard axis cannot solely be ascribed to a pinning of the domain walls at defects within the film during the reversal of the magnetisation.
Therefore, we suspect that our YAlIG thin films exhibit an additional magnetic anisotropy perpendicular to the film surface compensating the shape anisotropy. 

\subsection*{Ferromagnetic resonance}
\label{subsec:FMR}

\begin{figure}
\centering
\includegraphics{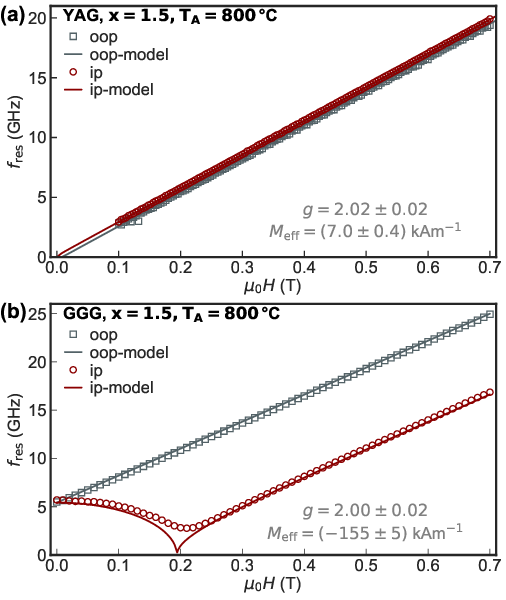}
\caption{\label{fig:FMR}FMR resonance $f_\mathrm{res} $ as a function of the applied magnetic field $H$ at a temperature of 200\,K for a YAlIG sample grown on YAG (a) and one sample grown on GGG (b). The modelled curve is based on the approach by Baselgia et al.\cite{Baselgia1988} and the total free energy density described in Eq. ~\ref{eq:YAlIG_FMR_Free_energy2}. The corresponding $M_\mathrm{eff}$ and $g$ are displayed in the plot.}
\end{figure}

The bb-FMR measurements demonstrate that the substrate-dependent epitaxial deformation can induce perpendicular magnetic anisotropy in the YAlIG thin films. 
The dynamic magnetic properties of our YAlIG thin films are comparable to those of other rare earth garnet thin films. Two representative samples with the same nominal chemical composition of $x=1.5$, which were both annealed at $T_\mathrm{A}=800$\,°C for 4h, but on different substrates have been used. The FMR resonance frequencies $f_\mathrm{res}$ as a function of magnetic field $H$ at 200\,K for the YAlIG thin film on YAG and GGG are shown in Fig.~\ref{fig:FMR}a and Fig.~\ref{fig:FMR}b, respectively. The measurement temperature of 200\,K has been chosen to be below the Curie temperature of YAlIG on YAG measured by SQUID VSM ($T_\mathrm{c}=262$\,K). The existence of the FMR signal confirms the ferrimagnetic nature of the YAlIG thin film.

To extract the g-factor $g$ and the effective magnetisation $M_\mathrm{eff}$, we model $f_\mathrm{res}$ using the approach of Baselgia et al.\cite{Baselgia1988} based on the total free energy density $F$ given by

\begin{equation}
\label{eq:YAlIG_FMR_Free_energy2}
\begin{split}
F =&-\mu_0M_\mathrm{s} H (sin\theta_{M}sin\theta_{H}cos(\phi_{M}-\phi_{H}) +cos\theta_{M}cos\theta_{H})\\&+\frac{1}{2}\mu_0M_\mathrm{s} M_\mathrm{eff}cos^2\theta_{M},
\end{split}
\end{equation}

where $\theta_M$, $\phi_M$ and $\theta_H$, $\phi_H$ denote the polar and azimuthal angles of the magnetisation and the magnetic field, respectively.  The first term in Eq.~\ref{eq:YAlIG_FMR_Free_energy2} describes the Zeeman energy and the second term the effective demagnetisation energy. The effective magnetisation $M_\mathrm{eff}=M_\mathrm{s}-H_\mathrm{u}$ in this model includes the shape anisotropy field (equal to $M_\mathrm{s}$) and a perpendicular uniaxial anisotropy field $H_\mathrm{u}$.

Fig.~\ref{fig:FMR}a and \ref{fig:FMR}b show that the modelled $f_\mathrm{res}$ values are in good agreement with the experimental data. For both substrates, we extract a g-factor of approximately 2 which is expected for YAlIG as the magnetic moment mostly originates of electron spins. On the YAG substrate, $M_\mathrm{eff}=7.0$\,kA/m which is smaller than the saturation magnetisation $M_\mathrm{s}$. We can therefore conclude that the shape anisotropy is partially compensated by a perpendicular anisotropy. This is comparable to the result of the SQUID VSM measurement, although measured at a different temperature. On the GGG substrate, $M_\mathrm{eff}$ is negative, indicating that the perpendicular magnetic anisotropy is the dominant magnetic anisotropy. Thus, it is possible to grow YAlIG on GGG as a PMA (perpendicular magnetic anisotropy) material. This large perpendicular magnetic anisotropy, in particular compared to the sample grown on YAG, is caused by a magneto-elastic anisotropy. As discussed in the XRD analysis, the YAlIG thin films grown on GGG are epitaxially strained so that the film is under tensile strain in the film plane. In this case, previous reports also suggest a strong perpendicular magneto-elastic contribution and the possibility of PMA in rare earth garnet thin films \cite{Ding2020,Soumah2018,Ciubotariu2019}. We would anticipate a similar correlation for our YAlIG thin films grown on GGG.

The dynamic properties of a YAlIG thin film are evaluated based on the extracted resonance linewidth $\Delta f$. At $f_\mathrm{res}=10$\,GHz, the frequency linewidth is $\Delta f=200\,$MHz for a YAlIG thin film grown on GGG, which corresponds to a magnetic field linewidth of $\Delta B=7.1\,$mT ($\Delta B=\frac{2 \pi \hbar}{g\mu_B}\Delta f$). This value is within the range reported for strained rare earth garnet thin films ($\Delta B=0.3-7.5\,$mT) \cite{Soumah2018,Lui2019,Ding2020,Ciubotariu2019} and close to the range reported for sputtered YIG thin films ($\Delta B=0.35-7\,$mT)\cite{Ding2020,Yamamoto2004,Bhoi2018,Chang2014}. The Gilbert damping $\alpha$ is determined by evaluating the frequency dependence of the extracted resonance linewidths (not shown). In our YAlIG thin films, $\alpha$ is of the order of $10^{-3}$, which is comparable to other sputtered garnet thin films \cite{Bhoi2018,Lui2019,Ciubotariu2019}, but still higher than the best values reported for highest quality YIG thin films \cite{Hauser2016}.

\section*{Conclusion}

We report the successful growth of ferrimagnetic Y$_{3}$Fe$_{5-x}$Al$_{x}$O$_{12}$ films via RF sputtering for substitution levels of $x=1.5,1.75,2$ on single crystalline YAG (111) and GGG (111) substrates. A post-annealing step in reduced oxygen atmosphere is necessary for the crystallisation of YAlIG. On YAG substrates with a higher lattice mismatch, relaxed YAlIG(111) films are obtained with a higher degree of defects. The lower lattice mismatch with the GGG substrates results in the growth of fully strained perpendicular compressed YAlIG(111) layers of higher crystalline quality as compared to the samples on YAG.
SQUID VSM reveals that the Curie temperature of the sputtered YAlIG thin films on YAG can be controlled by the Al$^{3+}$ substitution level. We achieve \tc values down to 102\,K, which is well below room temperature. We also show low coercivity comparable to other sputtered YIG thin films. 
Broadband-FMR measurements evidence a dominant perpendicular magnetic anisotropy in the YAlIG thin film grown on GGG caused by the epitaxial strain, whereas no such effect can be observed in the thin films grown on YAG. We also find that the FMR resonance linewidth and Gilbert damping are similar to other rare earth garnet thin films.
However the chemical composition of the YAlIG films, measured via RBS, is not equal to the nominal substitution level of the sputtering targets. This effect has to be considered for a quantitative control of the chemical composition and hence the magnetic properties of Al-substituted YIG.

\section*{Acknowledgements}

We thank Andy Thomas and Richard Schlitz for helpful discussions. We gratefully acknowledge the HZDR Ion Beam Centre for the RBS experiments and analysis. This work was supported by Deutsche Forschungsgemeinschaft (DFG, German Research Foundation) – Project-ID 425217212, SFB 1432 and the project GO 944/9-1.

\bibliography{YAlIG.bbl}

\appendix

\end{document}